\documentclass[twocolumn,showpacs,superscriptaddress,preprintnumbers,amsmath,amssymb]{revtex4-1}
\usepackage{graphicx}
\usepackage{dcolumn}
\usepackage{appendix}
\usepackage{bm}

\usepackage[usenames,dvipsnames]{xcolor}
\usepackage{subfigure}
\usepackage{bbm}
\usepackage{enumerate}
\usepackage[
  bookmarks=true,
  colorlinks,
  linkcolor=black,
  urlcolor=black,
  citecolor=black,
  plainpages=false,
  pdfpagelabels,
  final,
  breaklinks=true
]{hyperref}

\usepackage{titlesec}
\usepackage{array}
\usepackage{dsfont}
\usepackage{physics}

\begin{document}

\title{Entanglement and squeezing of the optical field modes in high harmonic generation}

\author{Philipp Stammer}
\email{philipp.stammer@icfo.eu}
\affiliation{ICFO -- Institut de Ciencies Fotoniques, The Barcelona Institute of Science and Technology, 08860 Castelldefels (Barcelona), Spain}

\author{Javier Rivera-Dean}
\affiliation{ICFO -- Institut de Ciencies Fotoniques, The Barcelona Institute of Science and Technology, 08860 Castelldefels (Barcelona), Spain}

\author{Andrew S. Maxwell}
\affiliation{Department of Physics and Astronomy, Aarhus University, DK-8000 Aarhus C, Denmark}

\author{Theocharis~Lamprou}
\affiliation{Foundation for Research and Technology-Hellas, Institute of Electronic Structure \& Laser, GR-70013 Heraklion (Crete), Greece}
\affiliation{Department of Physics, University of Crete, P.O. Box 2208, GR-70013 Heraklion (Crete), Greece}

\author{Javier~Arg\"uello-Luengo}
\affiliation{ICFO -- Institut de Ciencies Fotoniques, The Barcelona Institute of Science and Technology, 08860 Castelldefels (Barcelona), Spain}

\author{Paraskevas Tzallas}
\affiliation{Foundation for Research and Technology-Hellas, Institute of Electronic Structure \& Laser, GR-70013 Heraklion (Crete), Greece}
\affiliation{ELI-ALPS, ELI-Hu Non-Profit Ltd., Dugonics tér 13, H-6720 Szeged, Hungary}

\author{Marcelo F. Ciappina}
\affiliation{Department of Physics, Guangdong Technion - Israel Institute of Technology, 241 Daxue Road, Shantou, Guangdong, China, 515063}
\affiliation{Technion - Israel Institute of Technology, Haifa, 32000, Israel}
\affiliation{Guangdong Provincial Key Laboratory of Materials and Technologies for Energy Conversion, Guangdong Technion - Israel Institute of Technology, 241 Daxue Road, Shantou, Guangdong, China, 515063}

\author{Maciej Lewenstein}
\affiliation{ICFO -- Institut de Ciencies Fotoniques, The Barcelona Institute of Science and Technology, 08860 Castelldefels (Barcelona), Spain}
\affiliation{ICREA, Pg. Lluis Companys 23, ES-08010 Barcelona, Spain}

\date{\today}

\begin{abstract}

Squeezing of optical fields, used as a powerful resource for many applications, and the radiation properties in the process of high harmonic generation have thus far been considered separately. In this Letter, we want to clarify that the joint quantum state of all the optical field modes in the process of high harmonic generation is in general entangled and squeezed. We show that this is already the case in the simplest scenario of driving uncorrelated atoms by a classical laser light field. The previous observation of product coherent states after the high harmonic generation process is a consequence of the assumption that the ground state depletion can be neglected, which is related to vanishing dipole moment correlations. Furthermore, we analyze how the resulting quadrature squeezing in the fundamental laser mode after the interaction can be controlled and explicitly show that all field modes are entangled. 

\end{abstract}

\maketitle

\emph{Introduction.} -- 
The field of attosecond science \citep{krausz2009attosecond} has recently established connections to the field of quantum optics and quantum information science \cite{bhattacharya2023strong, lewenstein2022attosecond}. This is mainly due to the efforts in describing systems driven by strong light fields with fully quantized approaches \cite{stammer2023quantum}, going beyond the extremely successful (semi)-classical methods \cite{amini2019symphony}.
In particular, the process of high harmonic generation (HHG) \cite{lewenstein1994theory, corkum1993plasma} has been the subject of various investigations to gain novel insights into the radiation properties of the scattered field. These insights were elusive from a semi-classical perspective without the quantization of the electromagnetic field \cite{gorlach2020quantum, lewenstein2021generation, rivera2022strong}. It was shown, that the final quantum state of the field modes are given by product coherent states \cite{lewenstein2021generation, rivera2022strong}, which holds when assuming small depletion of the ground state and is related to vanishing dipole moment correlations \cite{sundaram1990high} (note that a coherent state driving field needs to be assumed as well \cite{stammer2023role}). However, further analysis has shown that the field state generated in the HHG process is entangled \cite{stammer2022high, stammer2022theory}. 
In this work, we show how the description of the final quantum optical state after HHG by means of product coherent states is a consequence of the aforementioned approximations \cite{stammer2023quantum2}. In particular, we show that all field modes are naturally entangled, and further show that each mode exhibits squeezing. As a proof of principle we show that the quadrature squeezing of the fundamental driving mode can be tuned by varying the carrier-envelope phase (CEP) of the driving laser field allowing the angle of the squeezed field quadrature to rotate. 
This is particularly interesting since the squeezing already occurs in the simplest scenario when driving uncorrelated atoms with classical laser light. Thus far, squeezed light has been considered to drive the HHG process \cite{gorlach2023high, stammer2023limitations}, showing an extension of the generated spectrum. However, the other field properties besides the spectrum, such as squeezing \cite{gorlach2020quantum} or quantum coherence \cite{stammer2023role}, only received very limited attention in HHG.
Here, we will observe two-mode squeezing between all field modes participating in the process, which naturally leads to an entangled state between all modes \cite{josse2004entanglement}.
Generating such squeezed and massive entangled states is of importance for modern quantum technologies \cite{braunstein2005quantum, adesso2007entanglement, acin2018quantum}. This further highlights the use of strong-laser driven systems for quantum state engineering of light \cite{stammer2023quantum, stammer2022theory}, which has already shown the ability to generate non-classical states of light by means of optical Schrödinger cat-states \cite{lewenstein2021generation} having sufficient photon numbers to induce non-linear phenomena \cite{lamprou2023nonlinear}.
This work further manifest the intrinsic properties in the HHG process beyond the semi-classical framework by showing that entanglement and squeezing naturally occurs even in simple gas targets.


\emph{Quantum optical HHG without dipole correlations.} -- 
In the quantum optical description of HHG the interaction of an intense laser field, described by the coherent state $\ket{\alpha}$, with an atom in the ground state $\ket{g}$ is given by the following interaction Hamiltonian (a detailed derivation can be found in \cite{stammer2023quantum})
\begin{align}
    H_I(t) = - {d}(t) {E}_Q(t),    
\end{align}
where the electric field operator 
\begin{align}
    {E}_Q(t) = - i g \sum_{q=1}^N \sqrt{q} \left( a_q^\dagger e^{i q \omega t} - a_q e^{- i q \omega t} \right),
\end{align}
with $q \ge 2$ being the harmonic field modes, is coupled to the time-dependent dipole moment operator
\begin{align}
    {d}(t) = U_{sc}^\dagger (t) \, {d} \, U_{sc}(t). 
\end{align}

The dipole moment is in the interaction picture of the reference frame $U_{sc} (t) = \mathcal{T} \exp [- i \int_{0}^t d\tau H_{sc}(\tau)]$, with respect to the semi-classical Hamiltonian of the electron $H_{sc}(t) = H_A - {d}{E}_{cl}(t)$. This semi-classical Hamiltonian is the same as the one traditionally considered in semi-classical HHG theory \cite{lewenstein1994theory}, where $H_A = {p}^2/{2} + V({r})$ is the pure electronic Hamiltonian, and 
\begin{align}
    {E}_{cl}(t) = \Tr[{E}_Q(t) \dyad{\alpha}] = i g (\alpha e^{- i \omega t} - \alpha^* e^{i \omega t}),    
\end{align}
is the classical part of the driving laser electric field. 
Solving the dynamics for the parametric process of HHG, in which the electron is found in the ground state after the end of the pulse, the dynamical equation for the field state conditioned on the electronic state $\ket{g}$ is given by 
\begin{align}
    i \partial_t \ket{\Phi(t)} = - \bra{g} {d}(t){E}_Q(t) \ket{\Psi(t)},
\end{align}
where $\ket{\Phi(t)} = \bra{g} \ket{\Psi(t)}$, with $\ket{\Psi(t)}$ the state of the total system. 
The general solution for the field state conditioned on the electronic ground state is given by \cite{stammer2022theory}
\begin{align}
    \ket{\Phi(t)} = \bra{g} U(t) \ket{g} \ket{\Phi_i} = K_{HHG} \ket{\Phi_i},
\end{align}
with the conditioned evolution operator
\begin{align}
\label{eq:kraus_exact}
    K_{HHG} = \bra{g} \mathcal{T} \exp \left[ i \int_{0}^t dt^\prime {d}(t^\prime) {E}_Q(t^\prime) \right] \ket{g},
\end{align}
which solely acts on the initial field state $\ket{\Phi_i} = \ket{\{ 0_q \}}$. 
To obtain the exact solution for the field we introduce the identity on the electronic subspace $\mathds{1} = \dyad{g} + \int d v \dyad{{v}}$ (for the sake of simplicity we use a $1d$ momentum representation) \cite{lewenstein2021generation, stammer2023quantum, rivera2022strong}, and in the spirit of the strong field approximation (SFA)\cite{amini2019symphony} we neglect continuum-continuum transitions such that we have to solve the following set of coupled equations
\begin{align}
\label{eq:dgl_coupled1}
    \partial_t \ket{\Phi(t)}  &= i {E}_Q(t) \bra{g} d(t) \ket{g} \ket{\Phi(t)} \\&
     \hspace{0.3cm}+ i \int dv {E}_Q(t) \bra{g} {d}(t) \ket{{v}} \ket{\Phi({v},t)},\nonumber \\
    \partial_t \ket{\Phi({v},t) }  &= i {E}_Q(t)  \bra{{v}} {d}(t) \ket{g} \ket{\Phi(t)}, 
\end{align}
where we have defined $\ket{\Phi(v,t)} = \bra{v} \ket{\Psi(t)}$ for the continuum state $\ket{v}$ of the electron with velocity $v$.
Solving the exact dynamics for $\ket{\Phi(t)}$ in \eqref{eq:kraus_exact} is a tedious task, and approximations based on the physical situation under consideration are necessary. 
For instance, we can assume that the depletion of the electronic ground state is small \cite{lewenstein1994theory}, such that the second term in \eqref{eq:dgl_coupled1} can be neglected since it is proportional to the total continuum state amplitude. The remaining equation can then be solved and is given by \cite{stammer2022theory} 
\begin{align}
\label{eq:Kraus_approx}
    K_{HHG} \simeq \mathcal{T} \exp \left[ i \int_{0}^t dt^\prime \expval{d(t^\prime)} E_Q(t^\prime) \right] = \mathbf{D}[\chi].
\end{align}
leading to a displacement operation for each mode $\mathbf{D}[\chi] = \prod_q D[\chi_q]$,
with the coherent state displacements $\chi_q \propto \int_0^t dt^\prime \expval{d(t^\prime)} e^{i \omega_q t^\prime}$, which are proportional to the Fourier transform of the dipole moment expectation value $\bra{g} d(t) \ket{g} = \expval{d(t)}$. Note that the approximation to obtain Eq.~\eqref{eq:Kraus_approx} is equivalent to neglecting dipole moment correlations of the electron \cite{sundaram1990high, stammer2022high, stammer2022theory}, such that the final field state
\begin{align}
\label{eq:HHG_final_product}
    \ket{\Phi (t)} = K_{HHG} \ket{\{ 0_q \} } = \ket{ \{ \chi_q \} },
\end{align}
is given by product coherent states. We shall now show that the actual field state in the process of HHG is not given by product coherent states when taking into account the dipole moment correlations, and we will see that the proper final field state is entangled and squeezed.

\emph{Including dipole moment correlations.} -- 
The crucial approximation to find the expression for the final field state in \eqref{eq:HHG_final_product} is based on the assumption of negligible depletion of the ground state amplitude, equivalent to neglecting dipole moment correlations of the electron. That is, the approximation from \eqref{eq:kraus_exact} to \eqref{eq:Kraus_approx} leading to the product coherent states of all field modes \eqref{eq:HHG_final_product}, due to a linear expression in the field operators $a_q^{(\dagger)}$ in \eqref{eq:Kraus_approx}. In typical experiments with moderate laser intensities ($\approx 1\times 10^{14}$ W/cm$^2$), this assumption is reasonable, but, if we increase the intensity to values around $\approx 5\times 10^{14}$ W/cm$^2$ or larger, the ground state depletion starts to play a role, depending on the atomic species under consideration. Theoretically speaking, the ground state depletion implies that the exact interaction Hamiltonian in the propagator \eqref{eq:kraus_exact} does not commute at different times and remains an operator in the total Hilbert space ($\mathcal{H_A} \otimes \mathcal{H}_F$) of the electron plus the field 
\begin{align}
    [H_I(t_1), H_I(t_2)] \in \mathcal{H_A} \otimes \mathcal{H}_F.    
\end{align}

In contrast, the commutator of the approximate interaction Hamiltonian $H_I^\prime(t) =  - \expval{d(t)} E_Q(t)$ is just a complex number, i.e. $[H_I^\prime(t_1) , H_I^\prime(t_2) ] \in \mathds{C}$, and thus, when solving \eqref{eq:kraus_exact}, the modes do not mix. Going beyond the linear term of the field operator ${E}_Q(t)$ would lead, for instance, to squeezing in the field modes. Furthermore, all field modes will become entangled due to the mixing of the field operators $a_q^{(\dagger)}$ of the different modes. 
We therefore, anticipate signatures of squeezing and entanglement between the modes when going beyond the approximation of neglecting dipole moment correlations. 
This is done by solving \eqref{eq:dgl_coupled1} without the assumption of negligible ground state depletion of the electron. We thus solve the dynamics for higher orders of the exact interaction Hamiltonian $H_I(t) = - d(t) E_Q(t)$ instead of $H_I^\prime = - \expval{d(t)} E_Q(t)$ as done in \eqref{eq:Kraus_approx}.
Especially, we want to find a solution of \eqref{eq:dgl_coupled1}, and have
\begin{align}
    \partial_t & \ket{\Phi(t)}  =   i {E}_Q(t) \expval{{d}(t)} \ket{\Phi(t)}  \\
    &  - E_Q(t) \int_0^t dt^\prime \int  dv  {d}^*_{vg}(t) {d}_{vg}(t^\prime) {E}_Q(t^\prime) \ket{\Phi(t^\prime)}, \nonumber
\end{align}
where $d_{vg}(t) = \bra{v} d(t) \ket{g}$.
In order to find the evolution operator for the state $\ket{\Phi(t)}$, we perform a Markov type approximation $\ket{\Phi(t^\prime)} \to \ket{\Phi(t)}$, and make use of the identity on the electronic subspace $\int dv \dyad{v} = I - \dyad{g}$, such that 
\begin{align}
    \partial_t \ket{\Phi(t)} &= \left[  i {E}_Q(t) \expval{{d}(t)} - \expval{\dot Q(t) Q(t)} \right] \ket{\Phi(t)}, 
\end{align}
where we have defined 
\begin{align}
    Q(t) = \int_0^t dt^\prime {E}_Q(t^\prime) \left[ {d}(t^\prime) - \expval{{d}(t^\prime)} \right]. 
\end{align}

Neglecting the commutators $[Q(t), Q(t^\prime)]$ we find~\footnote{The operator $Q(t)$ quantifies the deviation of the dipole moment from it's mean value, i.e. its fluctuations. And since the commutator $[Q(t),Q(t^\prime)]$ is related to the fluctuations of this operator, we neglect those fluctuations of the fluctuations.}
\begin{align}
    \expval{\dot Q(t) Q(t)} = \frac{1}{2} \pdv{t} \expval{Q^2(t)},
\end{align}
and thus, the evolution of the initial state reads
\begin{align}
\label{eq:derivation_factorizing}
    \ket{\Phi(t)} = \mathbf{D}[\chi] e^{- \frac{1}{2} \expval{Q^2(t)}} \ket{\Phi_i}.
\end{align}

Here, we have used that $[{E}_Q(t), Q(t^\prime)]$ is a complex number such that the two operations in \eqref{eq:derivation_factorizing} can be factorized.
The additional term is quadratic in $Q(t)$ and thus in the electric field operator and therefore leads to squeezing and mixing between the field modes. Evaluating this term we find 
\begin{align}
\label{eq:Q_quadratic_general}
    \expval{Q^2(t)} = & \int_0^t \int_0^t dt' dt'' {E}_Q(t') {E}_Q(t'') \\
    & \expval{\left[ {d}(t') - \expval{{d}(t')} \right] \left[ {d}(t'') - \expval{{d}(t'')} \right] }. \nonumber
\end{align}

As done in previous works \cite{lewenstein2021generation, rivera2022strong}, we will focus on the fundamental mode of the driving field (with operators $a_1^{(\dagger)} = a^{(\dagger)} $) since the amplitudes of the harmonics are much smaller. 
We can thus write 
\begin{align}
    \expval{Q^2}= & - g^2 \left[ (a^\dagger)^2 \expval{\mathcal{D}(\omega)\mathcal{D}(\omega) } + a^2 \expval{\mathcal{D}^\dagger(\omega) \mathcal{D}^\dagger(\omega)}   \right. \nonumber \\
    & \left. -  a^\dagger a  \expval{\mathcal{D}(\omega)\mathcal{D}^\dagger(\omega) } - a a^\dagger \expval{\mathcal{D}^\dagger(\omega) \mathcal{D}(\omega)} \right],
\end{align}
where we have considered $t \to \infty$ and defined 
\begin{align}
\label{eq:fluctuations_fundamentel}
    \mathcal{D}(\omega) = \int_0^\infty dt' e^{i \omega t'} \left[ {d}(t') - \expval{{d}(t')} \right].
\end{align}

We expect all the averages to be comparable, i.e. $\abs{\expval{\mathcal{D}(\omega)\mathcal{D}(\omega) }} = \expval{\mathcal{D}^\dagger(\omega) \mathcal{D}(\omega)}  = B,$
where we have used that $\mathcal{D}^\dagger(\omega) \mathcal{D}(\omega)$ is self-adjoint and defined $\expval{\mathcal{D}(\omega)\mathcal{D}(\omega) }^* = \expval{\mathcal{D}^\dagger(\omega)\mathcal{D}^\dagger(\omega) } \equiv B e^{- 2 i \psi}$,
such that we can write 
\begin{align}
    \expval{Q^2} = 2 g^2 B \left[ \frac{i}{\sqrt{2}} \left( a^\dagger e^{i \psi} - a e^{-i \psi} \right)\right]^2.
\end{align}

Interestingly, this includes a quadrature operator, leading to squeezing along the momentum quadrature 
\begin{align}
\label{eq:quadrature_operator}
    P_\psi = \frac{i}{\sqrt{2}} \left( a^\dagger e^{i \psi} - a e^{- i \psi} \right).
\end{align}

We can now write down the final field state of the fundamental mode after the process of HHG, which is given by (in the laboratory frame)
\begin{align}
\label{eq:state_fundamental_squeezed}
    \ket{\Phi_1} = {D}[\alpha] {D}[\chi_1] S(\psi) \ket{0},
\end{align}
which is a displaced and squeezed vacuum state, i.e. a high photon number squeezed coherent state, with the squeezing operator given by 
\begin{align}
    S(\psi) = \exp[- g^2 B P_\psi^2].   
\end{align}

We note that varying the squeezing phase $\psi$ in \eqref{eq:quadrature_operator} allows to consider squeezing along different directions given by the quadrature operator $P_\psi$.
\begin{figure}
    \centering
	\includegraphics[width=1\columnwidth]{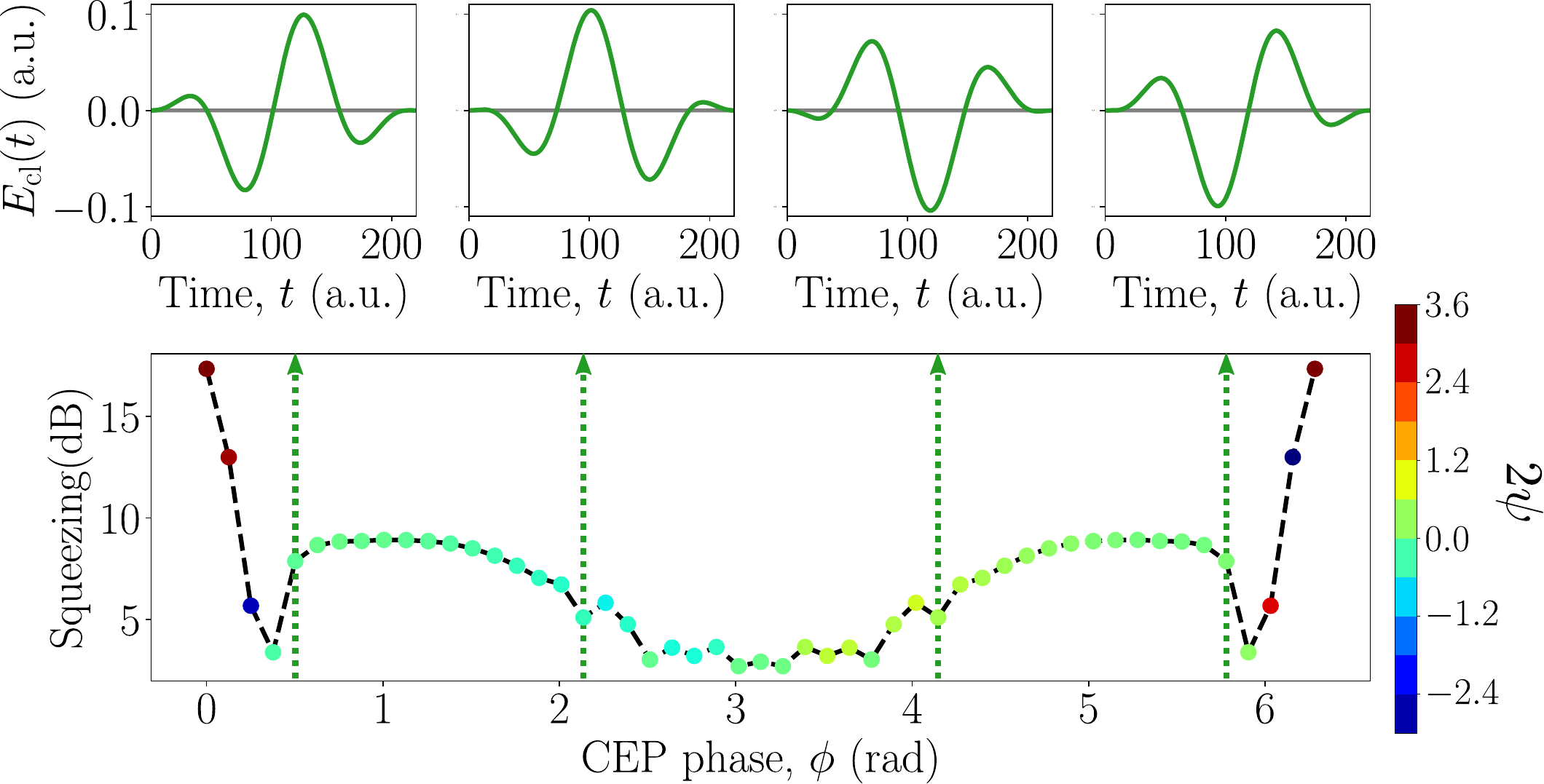}
	\caption{Squeezing parameter of the fundamental mode $\expval{\mathcal{D}(\omega)\mathcal{D}(\omega)}$ versus the carrier-envelope phase (CEP). To convert to dB units, we employed the relation $[\text{db}] = 10\log_{10}(e^{2\abs{r}^2})$, with $r \equiv - g^2 B N_{\text{at}}$ denoting the squeezing parameter. The driving laser field used in the calculations has an intensity of $4\times 10^{14} \text{W}/\text{cm}^2$ with a central frequency $\omega = 0.057$ a.u. (corresponding to a wavelength $\lambda = 800$ nm) and a sin$^2$-shaped envelope with 2 cycles of total duration (approximately $5$ fs). The target is an H atom in its ground, $1s$, state. Here, we have considered that $N_{\text{at}} = 5\times 10^{13}$ atoms independently contribute to HHG in the interaction region~\cite{lewenstein2021generation}. The laser field is plotted in the upper panels for each of the CEP values indicated. The arrows in this plot correspond to the values for which the Wigner function is shown in Fig.~\ref{fig:wigner}.}
  \label{fig:squeezing}
\end{figure}
In order to quantify the squeezing in the fundamental mode, we need the correlations of the dipole moment fluctuations around its mean $\mathcal{D}(\omega)$ in \eqref{eq:fluctuations_fundamentel}. We therefore numerically solve~\footnote{See Supplementary Material for details on the model and its numerical implementation.}
\begin{align}
    \expval{\mathcal{D}(\omega) \mathcal{D}(\omega)} & = \int_0^t dt' \int_0^t dt'' e^{i \omega t'} e^{i \omega t''}  \\ 
    & \times  \left[ \expval{{d}(t') {d}(t'')} - \expval{{d}(t')} \expval{{d}(t'')} \right], \nonumber
\end{align}
from which we obtain the single atom contribution to the squeezing $\abs {\expval{\mathcal{D}(\omega) \mathcal{D}(\omega)} } = B$. Taking into account that in the interaction region we have $N_{\text{at}}$ atoms independently contributing to the HHG process we can write the total squeezing power via the squeezing parameter $r \equiv -  g^2 B N_{\text{at}}$ in units of $[\text{db}] = 10\log_{10}(e^{2\abs{r}^2})$. 
In Fig.~\ref{fig:squeezing} we show the squeezing power in the fundamental mode after the HHG process for varying carrier-envelope phase (CEP) of the driving laser field. The squeezing phase $\psi = \frac{1}{2}\arg [\expval{\mathcal{D}(\omega) \mathcal{D}(\omega)}]$ is indicated with the color-coding for each CEP value. 
We observe that by varying the CEP we can control the squeezing power and the squeezing phase, which allows to rotate the angle of the quadrature operator along which the squeezing occurs. The arrows at different CEP values in Fig.~\ref{fig:squeezing} show different shapes of the electric field of the driving pulse. 
To further illustrate the effect of the squeezing on the fundamental mode, and to show the rotation of the quadrature operator, the Wigner function~\cite{royer_wigner_1977} of the state in \eqref{eq:state_fundamental_squeezed} is shown in  Fig. \ref{fig:wigner} for the CEP values indicated by the arrows in Fig.~\ref{fig:squeezing}. 
We can see significant squeezing along the field quadratures and the ability to rotate the squeezing ellipse by varying the CEP. Further, we can see how the symmetry between the pulse shapes (see Fig.~\ref{fig:squeezing}) is reflected in the orientation of the squeezing quadrature showing a rotation of the Wigner function in Fig.~\ref{fig:wigner} between (a) and (d) of $\pi$. The same holds between (b) and (c) in Fig.\ref{fig:wigner}, which emphasizes that there exists a relation between the CEP phase $\phi$ and the squeezing phase $\psi$, such that when applying a sign change and time-reversal symmetry operation on the driving pulse, the squeezing phase changes according to $\psi \to \pi - \psi$.

\begin{figure}
    \centering
	\includegraphics[width=1.0\columnwidth]{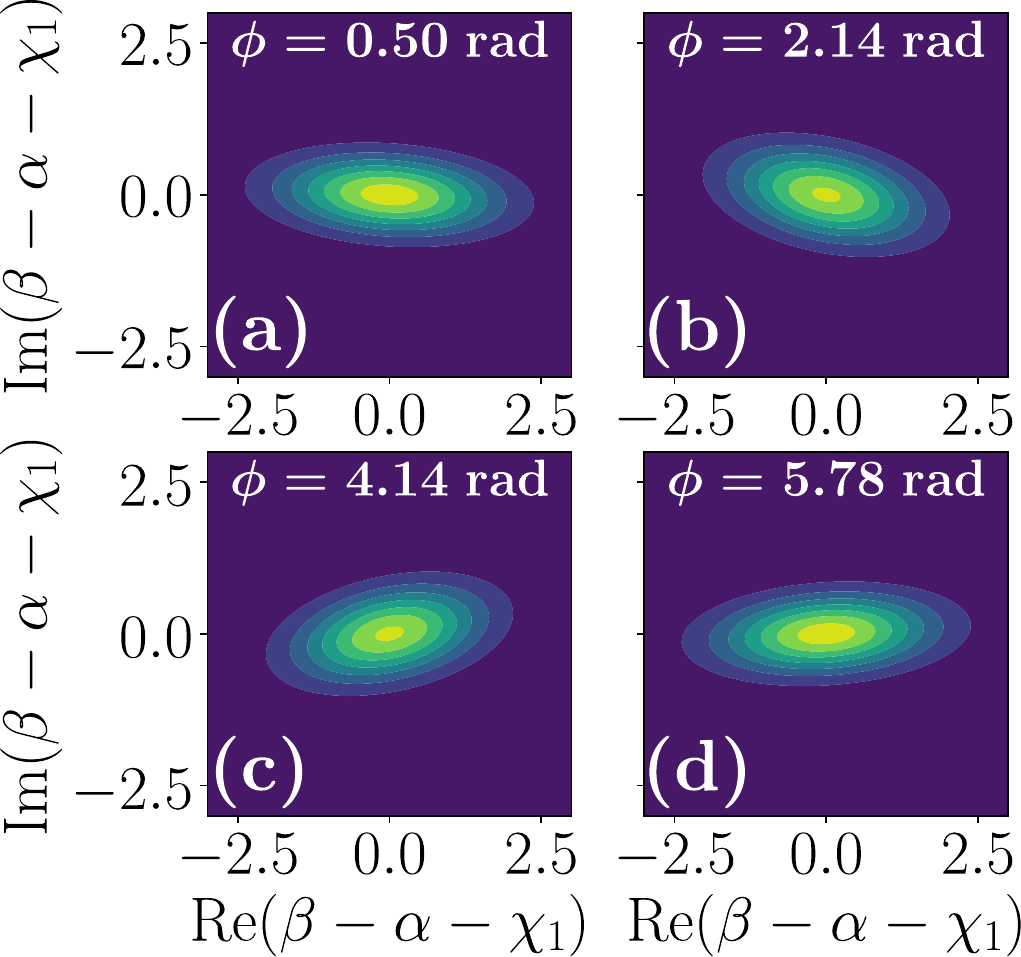}
	\caption{Wigner function $W(\beta)$ of the state of the fundamental mode \eqref{eq:state_fundamental_squeezed} for different values of the CEP $\phi$, with the same laser parameters as specified in the caption of Fig.~\ref{fig:squeezing}. The cases (a)-(d) correspond to the arrows indicated in Fig.~\ref{fig:squeezing}, illustrating the ability to control the rotation of the quadrature squeezing by varying the CEP.} 
      \label{fig:wigner}
\end{figure}

\emph{Entanglement in HHG.} -- 
Now, we shall take into account all harmonic field modes, to show that the final state of the field is in general entangled. Starting from \eqref{eq:Q_quadratic_general} by considering all modes in the field operator, we can write
\begin{align}
    \expval{Q^2} = & - g^2 \sum_{q,p} \sqrt{qp} \left[ a_q^\dagger a_p^\dagger \expval{\mathcal{D}(\omega_q) \mathcal{D}(\omega_p)} \right.  \\
    & \left. - a_q^\dagger a_p \expval{\mathcal{D}(\omega_q) \mathcal{D}^\dagger(\omega_p)} + \text{h.c.} \right]. \nonumber
\end{align}

This is a generic bilinear Hermitian operator for all the bosonic field modes participating in the process of HHG. The terms in which $q=p$, i.e.~the pairs of identical harmonic photons created or annihilated, correspond to a single mode squeezing, and the equivalent terms with $q\neq p$ correspond to two-mode squeezing. It is known that those two-mode squeezing terms are responsible for entanglement between the two modes \cite{josse2004entanglement}. This further manifests that the actual field state in the process of HHG is in general entangled. 
In the particular case of HHG, where many different field modes participate in the process, this two-mode squeezing leads to massive entangled states. In view of quantum information processing tasks with continuous variable systems \cite{braunstein2005quantum, adesso2007entanglement} the detection and characterisation of entanglement in continuous variable systems is of great importance \cite{duan2000inseparability, simon2000peres}. Taking into account that $N$ modes participate in the process of HHG, we expect the final field state to exhibit genuine multipartite continuous variable entanglement \cite{van2003detecting}, and its detection is subject of future investigation.

\emph{Conclusions.} -- 
We have shown that the final field state in the process of HHG is an entangled state between all field modes and that each mode is squeezed. The squeezing can be manipulated by varying the CEP of the driving field, which allows to rotate the quadrature operator along which the squeezing occurs. 
Showing that the field is entangled provides further usefulness of the recently emerging connection of attosecond physics with quantum information science \cite{lewenstein2022attosecond, bhattacharya2023strong}. 
Especially the generation of a genuine multipartite entangled continuous variable system is of fundamental and technological importance.
We anticipate that driving more complex systems \cite{pizzi2023light, rivera2022quantum, rivera2023bloch} or using non-classical driving fields \cite{gorlach2023high, even2023photon, stammer2023limitations} could lead to further insights into the entanglement and squeezing properties of the generated harmonic light.

\begin{acknowledgments}

P.S. acknowledges funding from the European Union’s Horizon 2020 research and innovation programme under the Marie Skłodowska-Curie grant agreement No 847517. 
J.R-D. acknowledges funding from the Government of Spain (Severo Ochoa CEX2019-000910-S and TRANQI), Fundació Cellex, Fundació Mir-Puig, Generalitat de Catalunya (CERCA program) and the ERC AdG CERQUTE. 
ICFO group acknowledges support from: ERC AdG NOQIA; MICIN/AEI (PGC2018-0910.13039/501100011033, CEX2019-000910-S/10.13039/501100011033, Plan National FIDEUA PID2019-106901GB-I00, FPI; MICIIN with funding from European Union NextGenerationEU (PRTR-C17.I1): QUANTERA MAQS PCI2019-111828-2); MCIN/AEI/ 10.13039/501100011033 and by the “European Union NextGeneration EU/PRTR" QUANTERA DYNAMITE PCI2022-132919 within the QuantERA II Programme that has received funding from the European Union’s Horizon 2020 research and innovation programme under Grant Agreement No 101017733Proyectos de I+D+I “Retos Colaboración” QUSPIN RTC2019-007196-7); Fundació Cellex; Fundació Mir-Puig; Generalitat de Catalunya (European Social Fund FEDER and CERCA program, AGAUR Grant No. 2021 SGR 01452, QuantumCAT \ U16-011424, co-funded by ERDF Operational Program of Catalonia 2014-2020); Barcelona Supercomputing Center MareNostrum (FI-2023-1-0013); EU (PASQuanS2.1, 101113690); EU Horizon 2020 FET-OPEN OPTOlogic (Grant No 899794); EU Horizon Europe Program (Grant Agreement 101080086 — NeQST), National Science Centre, Poland (Symfonia Grant No. 2016/20/W/ST4/00314); ICFO Internal “QuantumGaudi” project; European Union’s Horizon 2020 research and innovation program under the Marie-Skłodowska-Curie grant agreement No 101029393 (STREDCH) and No 847648 (“La Caixa” Junior Leaders fellowships ID100010434: LCF/BQ/PI19/11690013, LCF/BQ/PI20/11760031, LCF/BQ/PR20/11770012, LCF/BQ/PR21/11840013). Views and opinions expressed are, however, those of the author(s) only and do not necessarily reflect those of the European Union, European Commission, European Climate, Infrastructure and Environment Executive Agency (CINEA), nor any other granting authority. Neither the European Union nor any granting authority can be held responsible for them. 
M. F. C.~acknowledges financial support from the Guangdong Province Science and Technology Major Project (Future functional materials under extreme conditions - 2021B0301030005) and the Guangdong Natural Science Foundation (General Program project No. 2023A1515010871).
P. Tzallas group at FORTH acknowledges support from: LASERLABEUROPE V (H2020-EU.1.4.1.2 grant no.871124), the H2020 project IMPULSE (GA 871161) and ELI--ALPS. ELI--ALPS is supported by the European Union and co-financed by the European Regional Development Fund (GINOP Grant No. 2.3.6-15-2015-00001).

\end{acknowledgments}


\bibliography{literatur}
\bibliographystyle{apsrev4-2}



\end{document}